\documentclass[preprint]{revtex4-2}
\usepackage{graphicx,amsmath,natbib,hyperref,comment,color}
\usepackage[normalem]{ulem}
\usepackage[usenames,dvipsnames]{xcolor}
\hypersetup{colorlinks=true, citecolor=Plum, linkcolor=Plum,
urlcolor=Plum}

\begin{document}

\title{Black hole horizons must be veiled by photon spheres}
\author{Raúl Carballo-Rubio}
\email[Corresponding author: ]{raul@sdu.dk}
\affiliation{CP3-Origins, University of Southern Denmark, Campusvej 55, DK-5230 Odense M, Denmark}
\author{Astrid Eichhorn}
\email{eichhorn@cp3.sdu.dk}
\affiliation{CP3-Origins, University of Southern Denmark, Campusvej 55, DK-5230 Odense M, Denmark}
\date{27 March 2024}

\begin{abstract}
Horizons and bound photon orbits are defining features of black holes that translate into key features of black hole images. We present a purely geometric proof that spherically symmetric, isolated objects with horizons
in gravity theories with null-geodesic propagation of light must display bound photon orbits forming a photon sphere. Identifying the key elements of the proof, we articulate a simpler argument that carries over to more general situations with modified light propagation and implies the existence of equatorial spherical photon orbits in axisymmetric spacetimes with reflection symmetry. We conclude that the \emph{non-}observation of photon rings with very-large-baseline interferometry would be a very strong indication against a horizon, irrespective of whether or not the image shows a central brightness depression.
\\
\begin{center}
{\bf Honorable Mention in the Gravity Research Foundation 2024 Awards for Essays on Gravitation.}
\end{center}
\end{abstract}

\maketitle
\newpage

\section{Motivation}
In general relativity, vacuum black holes have horizons and photon shells (which suitably generalize photon spheres to rotating situations~\cite{Johnson:2019ljv}). Together, these make up the characteristic properties of black hole images in general relativity, namely a central brightness depression -- due to emission that is absorbed by the horizon~\cite{Bronzwaer:2021lzo} -- surrounded by a set of photon rings -- due to nearly-bound photon orbits that form lensed images of the accretion disk~\cite{Johnson:2019ljv,Bardeen:1973tla}.

Beyond general relativity, exotic compact objects may exist that have neither horizons nor photon shells~\cite{Eichhorn:2022oma}, or horizonless spacetimes with photon shells may arise~\cite{Carballo-Rubio:2023mvr}. Thus, a critical open question is whether the presence of a horizon necessarily implies the existence of a photon shell. This issue has received little attention in the literature, with an existing proof in spherical symmetry requiring constraints on matter fields in the form of energy conditions~\cite{Claudel:2000yi}. 

Here, we address this question by proving that, under a set of assumptions that invoke neither the equations of motion of general relativity, nor energy conditions, a spherically symmetric, asymptotically flat spacetime with an event horizon must also have a photon sphere. We then isolate the key geometric elements in the proof and discuss how these are still present in more general situations.

\section{Geometric proof in spherical symmetry}

A general parametrization of spherically symmetric, stationary spacetimes is provided by the line element
\begin{equation}\label{eq:metric}
\text{d}s^2=g_{vv}(r)\text{d}v^2+2g_{vr}(r)\text{d}r\text{d}v+r^2\left(\text{d}\theta^2+\sin^2\theta\,\text{d}\varphi^2\right),
\end{equation}
in which the geodesic equation can be written as
\begin{equation}\label{eq:gen_ngeo}
\frac{d^2 x^\mu}{dv^2}+\Gamma^\mu_{\alpha\beta}\frac{dx^\alpha}{dv}\frac{dx^\beta}{dv}=0.
\end{equation}
Without loss of generality, we can focus on equatorial circular null geodesics with radius $r=r_{\rm ps}$ and tangent vector
\begin{equation}\label{eq:null_geo}
\frac{dx^\mu}{dv}=\left(1,\frac{dr}{dv},\frac{d\theta}{dv},\frac{d\varphi}{dv}\right)=\left(1,0,0,\dot{\varphi}\right).
\end{equation}
These equations suffice to construct the proof we are looking for:
\begin{itemize}
\item From $\text{d}s^2=0$ being satisfied by null geodesics and Eq.~\eqref{eq:metric}, we obtain
\begin{equation}\label{eq:res1}
\sin^2\theta\,\dot{\varphi}^2=-\frac{g_{vv}(r)}{r^2}.
\end{equation}
\item From $\text{d}^2 r/\text{d}v^2=0$, which is a corollary of $\text{d} r/\text{d}v=0$ in Eq.~\eqref{eq:null_geo}, and Eq.~\eqref{eq:gen_ngeo}, we obtain
\begin{equation}\label{eq:res2}
\sin^2\theta\,\dot{\varphi}^2=-\frac{
g'_{vv}(r)}{2r}.
\end{equation}
The only two Christoffel symbols needed for this result are~\cite{Burtscher:2014msa}:
\begin{equation}
\Gamma^r_{vv}=\frac{1}{2}\frac{g_{vv}g'_{vv}}{g_{vr}^2},\qquad \Gamma^r_{\varphi\varphi}=\frac{g_{vv}}{g_{vr}^2}r\sin^2\theta.
\end{equation}
\end{itemize}
Comparing Eqs.~\eqref{eq:res1} and \eqref{eq:res2}, we obtain the following condition defining the photon sphere:
\begin{equation}\label{eq:ps_cond}
g_{vv}(r_{\rm ps})=\frac{r_{\rm ps}}{2}g'_{vv}(r_{\rm ps}).
\end{equation}
For the Schwarzschild metric, Eq.~\eqref{eq:ps_cond} yields the standard result $r_{\rm ps}=3M$.

Eq.~\eqref{eq:ps_cond} allows us to show that any asymptotically flat spacetime with a horizon must contain a photon sphere. We have the following limiting behaviors:
\begin{itemize}
\item For $r\rightarrow\infty$, asymptotic flatness yields the constraints
\begin{equation}
\lim_{r\rightarrow\infty}g_{vv}(r)=1,\qquad \lim_{r\rightarrow\infty}g'_{vv}(r)=0.
\end{equation}
\item For $r\rightarrow r_{\rm h}$, the existence of a horizon yields the constraint
\begin{equation}
\lim_{r\rightarrow r_{\rm h}}g_{vv}(r)=0.
\end{equation}
\end{itemize}
Hence, starting from $r\rightarrow\infty$ and moving towards $r=r_{\rm h}$, the left-hand side of Eq.~\eqref{eq:ps_cond} starts being 1, while the right-hand side starts being 0. When reaching $r=r_{\rm h}$, the left-hand side vanishes. If $g_{vv}(r)$ is monotonically increasing and thus the derivative $g'_{vv}(r)$ is positive, then Eq.~\eqref{eq:ps_cond} is satisfied for some value $r_{\rm ps}\in[r_{\rm h},\infty)$. The same reasoning can be applied if the function $g_{vv}(r)$ is not monotonically
increasing, by noticing that there must exist a point $r=r_\star$ at
which there is a local maximum of the function, so that $g_{vv}(r)$ is monotonically
increasing in the interval $(r_{\rm h},r_{\star})$. It follows that
\begin{equation}
g_{vv}(r_{\star})\geq\frac{r_{\star}}{2}g'_{vv}(r_{\star})=0,
\end{equation}
while 
\begin{equation}
0=g_{vv}(r_{\rm h})\leq\frac{r_{\rm h}}{2}g'_{vv}(r_{\rm h}),
\end{equation}
thus implying that Eq.~\eqref{eq:ps_cond} must be satisfied in the interval $(r_{\rm h},r_{\rm ps})$. 

Our proof shows that Eq.~\eqref{eq:ps_cond} must be satisfied at least once, but it does not determine whether this can be satisfied at several distinct radii, so that one horizon comes with a collection of photon spheres. This situation has been discussed previously for specific models~\cite{Gan:2021pwu}.

There could be a limiting case in which the horizon and the photon sphere are coincident. This requires that
\begin{equation}
g_{vv}(r_{\rm h})=g'_{vv}(r_{\rm h})=0.
\end{equation}
It has previously been discussed that, for spherically symmetric black holes, the photon sphere coincides with the horizon in the extremal limit~\cite{Pradhan:2010ws,Jia:2017nen}. We have shown the converse statement: any spacetime with coincident horizons and photon spheres describes an extremal black hole, because extremality is signalled by a degenerate zero of $g_{vv}(r_h)$.

\section{Light bending and capture by rotating horizons}

The proof above relies on continuity properties of the metric functions which, together with boundary conditions at the horizon and the asymptotically flat region, imply that a photon sphere must exist in between. 

Instead of working with metric functions, we can construct a similar proof using a quantity with direct physical meaning by introducing a point-like source emitting light isotropically in its local orthonormal frame. These light rays either escape to infinity, are captured by the gravitational field of the black hole, or -- if a photon sphere exists -- stay forever at a finite radial distance to the black hole horizon. To show that a photon sphere exists,
we define the fraction of light rays that are captured by the black hole as a function of the position of the source $R$,
\begin{equation}
\Delta(R)\in[0,1].
\end{equation}
For instance, for a Schwarzschild black hole with mass $M$ (and using natural units $G=c=1$), this quantity is given by~\cite{Carballo-Rubio:2018jzw}
\begin{equation}
\Delta_{\rm Sch}(R)=\frac{1}{2}\left\{1-\left(1-\frac{3M}{R} \right)\sqrt{1+\frac{6M}{R}}\right\}.
\end{equation}
The boundary conditions for this quantity are:
\begin{itemize}
\item For $R\rightarrow\infty$, asymptotic flatness yields the constraint
\begin{equation}
\lim_{R\rightarrow \infty}\Delta(R)=0.
\end{equation}
\item For $R\rightarrow r_{\rm h}$, the existence of a horizon yields the constraint
\begin{equation}
\lim_{R\rightarrow r_{\rm h}}\Delta(R)=1.
\end{equation}
\end{itemize}
\begin{figure}
    \centering
\includegraphics[width=0.7\linewidth]{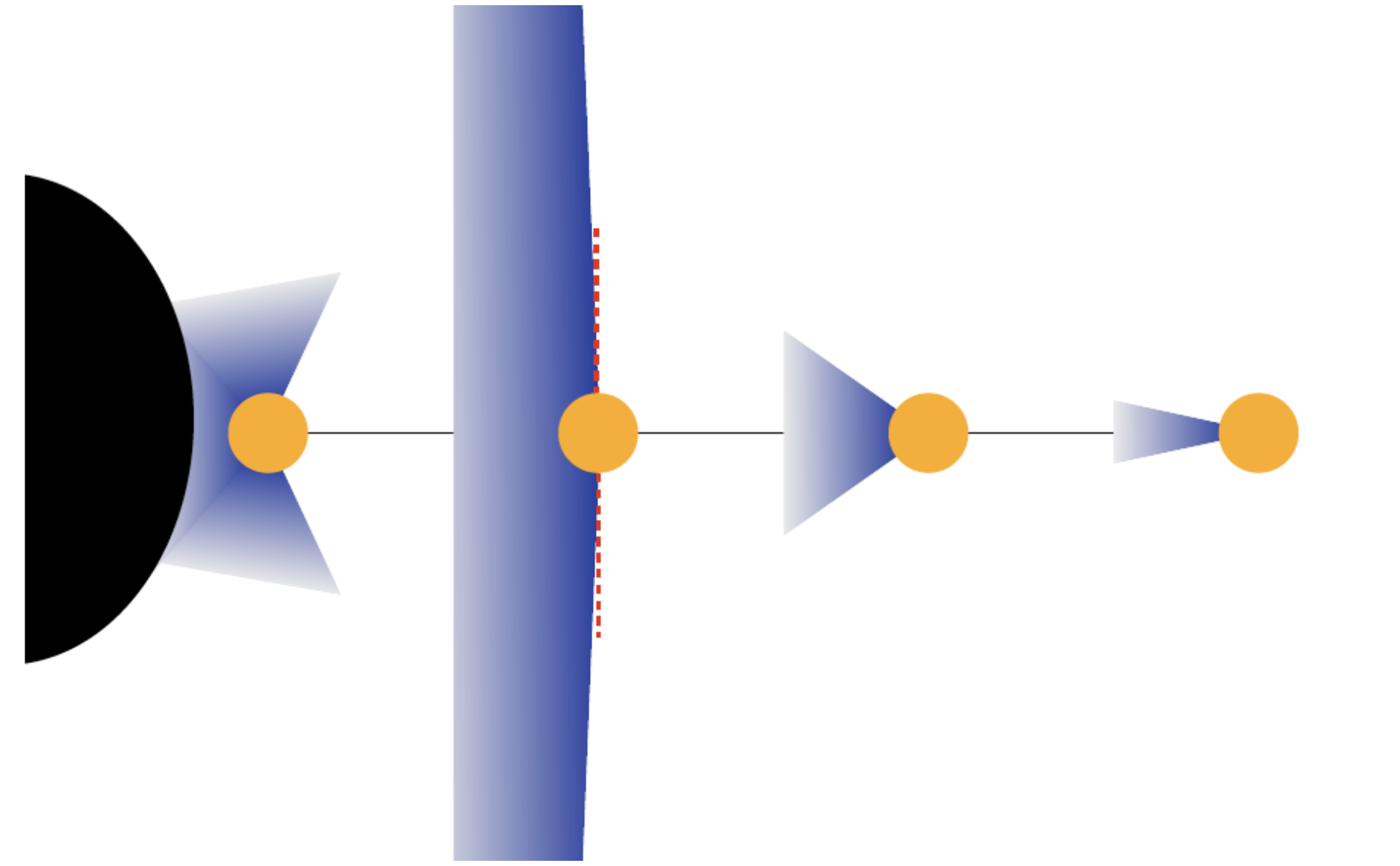}
    \caption{Illustration of the main idea for the proof of the existence of a photon sphere. The yellow circles indicate sources of light that emit isotropically. The solid angle from which light rays are captured by the horizon is indicated by the blue shaded areas. For a special radius, exactly half of the emission is captured, such that the transition between capture and escape happens at an angle of $\pi/2$. Thus, a  circular orbit exists, on which light forever circles the black hole at constant radius (note that this orbit is marginally stable, which cannot be deduced from our proof).}
    \label{fig:1}
\end{figure}
Continuity of $\Delta(R)$ implies that there exists a radius $R=r_{\rm ps}$ for which
\begin{equation}
\Delta(r_{\rm ps})=\frac{1}{2}.
\end{equation}
As we have anticipated in the notation, the specific value taken by $\Delta(r_{\rm ps})$ indicates the existence of a photon sphere. The argument is as follows (see also Fig.~\ref{fig:1}). For any value of $R$, there is always a critical emission angle $\hat{\varphi}=\pi/2$ (defined in the local orthonormal frame of the source or, equivalently, with respect to the radial line connecting the source to $r=0$), such that trajectories emitted at smaller angles are captured and trajectories at larger angles escape. By definition,  exactly half of the emission from the source is captured and half escapes if and only if this angle is $\pi/2$. In this situation, a marginally unstable orbit at $r_{\rm ps}$ exists, i.e., there is a third, special type of trajectory, which neither falls into the black hole nor escapes, but instead circles the black hole forever at constant radius, which signals the existence of a photon sphere. 

The above  light-bending argument does not make use of the null geodesic equation, which is an important ingredient of our geometric proof for spherically symmetric spacetimes. Going beyond general relativity, there is, however, no reason to expect that the null geodesic equation continues to describe the propagation of light. For instance, non-minimal couplings between photons and spacetime curvature, arising even in the leading-order quantum effects in quantum electrodynamics~\cite{Bastianelli:2008cu}, can lead to deviations. We anticipate that our proof carries over to such cases. Accounting for the modified propagation of light, the location of the event horizon will no longer be given by $g_{vv}=0$ and Eq.~\eqref{eq:gen_ngeo} will change by a related term. We thus anticipate that a similar proof to the one we presented can be formulated. Based on our light-bending argument, the modifications should not spoil the relationship between horizons and photon spheres in spherical symmetry.

\begin{figure}[h!]
    \centering
\includegraphics[width=0.35\linewidth]{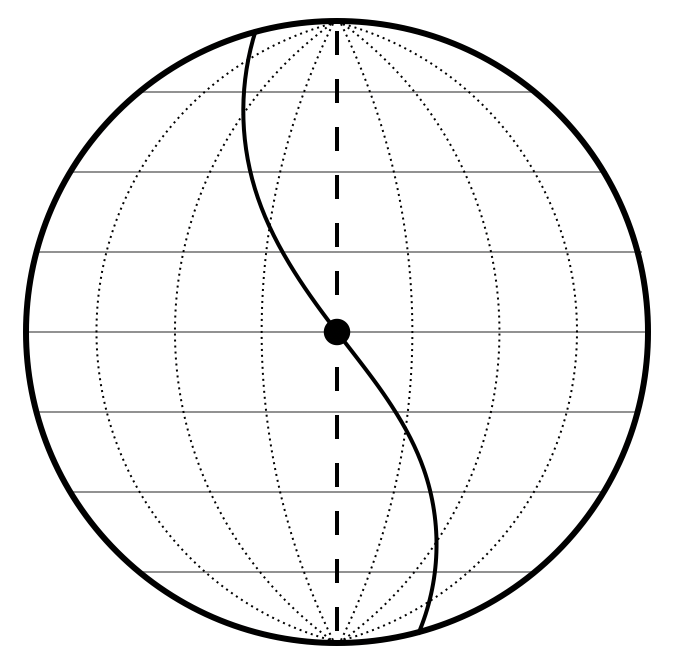}
    \caption{Hemisphere generated by prograde (retrograde) light rays emitted by the point-like source at $(R,\theta_0,\varphi_0)$. Eq.~\eqref{eq:delta_1/2} being satisfied implies that the vertical dashed line of $\hat{\varphi}=\pi/2$ for prograde ($\hat{\varphi}=3\pi/2$ for retrograde) is crossed at least once, thus signalling the presence of an initially spherical motion.}
    \label{fig:2}
\end{figure}

\newpage
Some elements in this proof, based on the behavior of light rays emitted by a point-like source, can be generalized to axisymmetric spacetimes. The lack of time reversal symmetry makes it necessary to distinguish between the prograde and retrograde light rays emitted by the source, the position of which can be specified in suitable coordinates as $(R,\theta_0,\varphi_0)$:
\begin{equation}
\Delta_\pm(R,\theta_0)\in[0,1],
\end{equation}
with $+$ for prograde light rays and $-$ for retrograde light rays. Both fractions satisfy the same boundary conditions as in the spherically symmetric case for $R\rightarrow\infty$ and $R=r_{\rm h}(\theta_0)$. Continuity implies that there exists hypersurfaces $R=r^\pm_{\rm ps}(\theta_0)$ in which
\begin{equation}\label{eq:delta_1/2}
\Delta_\pm(r^\pm_{\rm ps}(\theta_0),\theta_0)=\frac{1}{2}.
\end{equation}
Extracting the implications of the equation above is more subtle than in the spherically symmetric case. In the spherically symmetric case, it implied the existence of several light rays with $\hat{\varphi}=\pi/2$ (or $\hat{\varphi}=3\pi/2$) that are not captured nor escape to infinity and are therefore bound. This is a consequence of the lack of dependence on the polar angle $\hat{\theta}$ in the orthonormal frame
in spherical symmetry, which is no longer guaranteed in axisymmetric situations. When taking into account the dependence on $\hat{\theta}$, it follows that there exists at least one light ray with $\hat{\varphi}=\pi/2$ for prograde (or $\hat{\varphi}=3\pi/2$ for retrograde), and some 
specific value of $\hat{\theta}$, that is bound and tangent to the sphere $r=R$ at $(\theta_0,\varphi_0)$.

The behavior of $\Delta_\pm(R,\theta_0)$ does not seem to have enough information to determine, in general, whether these light rays stay tangent to the sphere $r=R$ and therefore lead to spherical orbits. However, in the specific case in which $\theta_0=\pi/2$, additional symmetries suffice to constrain these bound light orbits to be spherical. A reflection symmetry around this plane implies the lack of dependence on the angle $\hat{\theta}$ in the plot in Fig.~\ref{fig:2}, implying in particular that light rays with $(\hat{\theta},\hat{\varphi})=(0,\pi/2)$ and $(\hat{\theta},\hat{\varphi})=(0,3\pi/2)$, respectively prograde and retrograde, are bound.
The circular symmetry of the equatorial plane is the last piece needed to show that these bound orbits are spherical.

\section{Falsifying the existence of horizons by looking for photon rings}

We have presented a simple proof that any spherically symmetric spacetime that is asymptotically flat and has a horizon must contain a photon sphere, a spherical photon shell. The proof is purely geometrical and 
provides a critical link to observations, because horizons are arguably not \emph{directly} observable~\cite{Abramowicz:2002vt}, though their existence can be favored by hybrid arguments mixing observations and theoretical constraints (see~\cite{Carballo-Rubio:2018vin,Carballo-Rubio:2023fjj} for up-to-date discussions). However, photon shells can be directly probed by very-long-baseline interferometry, see \cite{Broderick:2022tfu} for a first result in this direction using data from the Event Horizon Telescope.
 
In the presence of a photon shell, light emitted by the accretion disk of a supermassive black hole can travel directly towards the observer, or circle the supermassive black hole a number of times before escaping and reaching the observer. This leads to a ring structure in the image, collectively called the photon ring. This ring structure is composed of a tower of higher-order images of the accretion disk that have been distorted due to photons traveling and circling around the central object.

Hence, the lack of observational evidence for photon rings would be a very strong indication against a horizon. We stress that the converse is not true, i.e., detecting photon rings does not imply the existence of horizons, both because there exist objects with photon spheres but without horizons, but also because compact enough objects without photon spheres genericallly display dimmer photon rings~\cite{Carballo-Rubio:2022aed}. 

Precision tests of the shape and luminosity of photon rings therefore provide further opportunities to falsify both the existence of photon shells and horizons. With the planned next-generation Event Horizon Telescope~\cite{Tiede:2022grp,Ayzenberg:2023hfw} and proposals of deploying space-based stations~\cite{Himwich:2020msm,Gralla:2020srx} providing enhanced prospects for detecting photon rings, it is likely that the next years will be critical for our understanding of black holes.

\acknowledgments
This work is supported by a grant from VILLUM Fonden, no.~29405. A.~Eichhorn also acknowledges support by the Deutsche Forschungsgemeinschaft (DFG) under Grant No 406116891 within the Research Training Group RTG 2522/1.

\newpage
\bibliographystyle{apsrev4-1}
\bibliography{refs}

\end{document}